\documentstyle[aps,prl] {revtex}

\begin{document}

\title {Lyapunov instability and finite size effects 
in a system with long-range forces}

\author{   Vito Latora $^{(a)}$   }

\address{Center for Theoretical 
Physics, Laboratory for Nuclear Sciences 
and Department of Physics, 
\\ Massachusetts Institute 
of Technology, Cambridge, Massachusetts 02139, USA}

\author{  Andrea Rapisarda $^{(b)}$ }

\address{ Istituto Nazionale 
di Fisica Nucleare, Sezione di Catania
and Dipartimento di Fisica, \\   Universit\'a di Catania,
Corso Italia 57, I-95129 Catania, Italy } 

\author{   Stefano Ruffo  $^{(c)}$  }

\address{ Centro Internacional 
de Ciencias, Cuernavaca, Morelos, Mexico }

\date{July 28, 1997 -  revised version October 28, 1997\\
{\it accepted for publication in Physical Review Letter}}
\maketitle

\begin{abstract} 

We study the largest Lyapunov exponent $\lambda$ and the finite size effects of  a 
system of N fully-coupled classical particles, which shows a 
second order phase  transition.
Slightly below the critical energy density $U_c$, $\lambda$ shows a peak which persists 
for very large $N$-values ($N=20000$). We show, both numerically and analytically, 
that chaoticity is strongly related to kinetic energy fluctuations.
In the limit of small energy,  $\lambda$ goes to zero with a $N$-independent
power law: $\lambda \sim \sqrt{U}$. 
In the continuum limit the system is integrable in
the whole high temperature phase. 
More precisely, the behavior $\lambda \sim N^{-1/3}$ 
is found numerically for $U > U_c$ and justified on the basis of a random matrix 
approximation.

\end{abstract}


%%%%%%%%%%%%%%%%%%%%%%%%%%%%%%%%%%%%%%%%%
% comandi per prl in due colonne 
%\narrowtext
%%%%%%%%%%%%%%%%%%%%%%%%%%%%%%%%%%%%%%%%%%%%%%%%%%%%%%%%%%%

Recently, the  interest in phase transitions  occurring in finite-size 
systems and the study of the related dynamical features has stimulated 
the investigation of the so far obscure relation between macroscopic 
thermodynamical properties and microscopic dynamical ones. In this 
respect several papers appeared in the recent literature  in various 
fields ranging from solid state physics 
\cite{butera,berry,nayak,dellago,ruffo,yama,lapo1}
to lattice field theory~\cite{lapo3} and 
nuclear physics \cite{ata,cmd}, where there is presently a lively 
debate on multifragmentation phase transition~\cite{ata,cmd,bond,gsi,eos}. 
The general expectation is that there is a close connection between 
the increase of fluctuations at a phase transition and a rapid increase 
of chaoticity at the microscopic level.
In several pioneering papers a different behavior 
of the Largest Lyapunov Exponent (LLE) $\lambda$ was found,  
according to the order of the transition~\cite{nayak,lapo1,lapo3,cmd}.
In particular, a well pronounced peak in LLE has been found for
second order phase transitions,
while a sharp increase has been seen  for first order phase transitions.  
In the former case some universal features have also been found, 
i.e. different systems show the same behavior when properly scaled~\cite{cmd}.
In order to connect dynamical properties of systems of size $N$
to bulk phase transitions one has to explore the  continuum limit, $N \to \infty$.
This unfortunately is not always possible due to computer time limitations, 
and has been done very rarely.  In this  letter we present numerical 
investigations of the $N$-dependence of the LLE up to N=20000, a size for 
which we already observe a certain convergence to the continuum limit.
We have investigated a toy model consisting of N classical particles
moving on the unit circle and interacting via long-range forces \cite{ruffo}.
This model shows a second order phase transition from a clustered phase to 
a homogeneous one at $U_c=(E/N)_c=0.75$~\cite{ruffo}. Some results for the 
LLE of systems of moderate sizes ($N \approx 100$) have already been 
published~\cite{yama}.
The model, though relatively simple, has very general properties which 
enable us to explore the connections between phase transitions and 
dynamical features in finite systems. In particular, it could be relevant
for nuclear multifragmentation where one has 100-200 particles interacting
via long-range (nuclear and  Coulomb) forces \cite{bond}. In this latter case a 
very similar caloric 
curve has been observed \cite{gsi} and critical exponents
have been measured experimentally \cite{eos}.   

The main results of this letter are:

\noindent
1) The system is strongly chaotic just below the canonical transition energy 
$U_c$. The peak in $\lambda (U)$, found in~\cite{yama} for small systems, 
persists as $N\to \infty $. 
 
\noindent
2)  The increase of the LLE is related to the increase of kinetic energy 
fluctuations.

\noindent
3)  For $U \to 0$, $\lambda \to 0$ as $U^{\alpha}$ where the 
exponent is found to be $\alpha=0.5$.
Essentially no dependence on the system size is observed in this regime.
A similar result was found for other systems \cite{cmd,aldoprep}. 

\noindent
4) For $U >  U_c$, $\lambda \to 0$ as $N^{-{1\over 3}}$. This behavior 
is explained by means of a random matrix approximation~\cite{pari}.

\noindent
5) Long-living quasi-stationary states are found in the critical 
region. These states look very similar to those recently obtained in 
~\cite{ppp} and simulate a discrepancy between the 
canonical and the microcanonical ensemble very similar to that one 
found   
in refs.\cite{thir} and more recently by other 
authors~\cite{compa,laba,gross,tor}. The fact that they appear
near a second order phase transition might be related to
critical slowing down.

The Hamiltonian we consider is the following
\begin{equation}
H(q,p) = K + V~, 
\end{equation}
\noindent
where 
\begin{equation}
K=\sum_{i=1}^N {1\over2} p_i^2,  ~ ~ ~
V = {\epsilon\over 2N} \sum_{i,j=1}^N [ {1-cos(q_i-q_j)} ]
\end{equation}
\noindent
are the kinetic and potential energies. The model describes the motion of N
particles on the unit circle: each particle interacts with all the others.
One can define a spin vector associated with each particle 
$
{\bf m}_i = ( cos(q_i), sin(q_i) ) ~.
$
\noindent
The Hamiltonian then describes N classical spins similarly to the XY model, and a 
ferromagnetic or an antiferromagnetic behavior according to the positive or negative
sign of $\epsilon$ respectively ~\cite{ruffo}. In the following we will consider 
only the ferromagnetic (attractive) case and in particular  $\epsilon=1$. Results concerning 
the case $\epsilon=-1$ will be discussed elsewhere~\cite{latora}.
The order parameter is the magnetization ${\bf M}$, defined as
$
{\bf M} =   {1\over N} \sum_{i=1}^N {\bf m}_i = (M_x, M_y)~. 
$
It is convenient to rewrite the potential energy $V$ as 
\begin{equation}
V =   {N\over 2} (1 -  (M_x^2 + M_y^2))  ~= 
      {N\over 2} (1 -  M^2 )            ~.
\end{equation}
The equations of motion can then be written as
\begin{equation}
{d\over dt} q_i  =   p_i ~~~~,~~~~
{d\over dt} p_i  = -sin(q_i) M_x   + cos(q_i) M_y ~~~.
\label{eqmoto} 
\end{equation}
In order to calculate the LLE one must consider the limit 
\begin{equation}
{\lambda}   = \lim_{t\to \infty}  {1\over t} ln { d(t)\over d(0) } 
\end{equation}
with 
$d(t) =  \sqrt{ \sum_{i=1}^N  (\delta q_i)^2  + (\delta p_i)^2 }
$
the metric distance calculated from the infinitesimal displacements at time t. 
Therefore, one has to integrate along the reference orbit the linearized 
equations of motion 
\begin{equation}
\label{lin1}
{d\over dt} \delta q_i  =   \delta p_i ~~~~,~~~~ 
{d\over dt} \delta p_i  
~= -\sum_j {{\partial^2V}\over {\partial q_i \partial q_j} } \delta q_j~~,  
\end{equation}
\noindent
where the diagonal and off-diagonal terms are 
\begin{equation}
\label{der1}
{\partial^2V\over{\partial q_i^2}}
=   cos(q_i) M_x   + sin(q_i) M_y  - {1\over{N}}
\end{equation}
\begin{equation}
\label{der2}
{{\partial^2V}\over {\partial q_i \partial q_j} }
~= -{1\over{N}}cos(q_i-q_j)~~~,    ~~~i \neq j
~~.
\end{equation}
\noindent
Expression (\ref{der1}) can also be written for convenience as: 
\begin{equation} 
\label{diag}
{\partial^2V\over{\partial q_i^2}}
=M cos(q_i-\Phi) - \frac{1}{N} 
\end{equation}
\noindent
where $\Phi$ is the phase of {\bf M }.
We have integrated Eqs. (\ref{eqmoto}), (\ref{lin1}) using fourth order 
symplectic algorithms~\cite{yo} with a time step $\Delta t = 0.2$, adjusted
to keep the error in energy conservation below ${\Delta E\over E} =10^{-5}$.
The LLE was calculated by the standard method of Benettin et al.\cite{ben}.
The average number of time steps in order to get a good convergence was 
of the order $10^6$. We discuss in the following numerical results for system 
sizes in between N=100 and N=20000. 

In fig.~1 we plot the caloric curve, i.e. the temperature as 
a function of U, and we compare it with the theoretical canonical 
prediction~\cite{ruffo} (in the inset we show the magnetization).
Simulations performed starting from equilibrated initial
data, which are Gaussian in momenta at the given canonical temperature,
agree very well with canonical predictions.
In fact, it is possible to solve the stationary Vlasov equation, which 
represents the system in the $N \to \infty$ limit, and obtain, under the 
factorization hypothesis for the probability distribution, $P(q,p)=f(q)g(p)$,
and assuming $g(p)$ to be Gaussian
\begin{eqnarray*}
f(q) &=& \frac{1}{2 \pi I_0 (M/T)} \exp(\frac{M \cos (q -\Phi)}{T}) \\
g(p) &=& \frac{1}{\sqrt{2 \pi T}}\exp (-\frac{p^2}{2T})~.
\end{eqnarray*}
In the latter, $I_0$ is the modified Bessel function of zero order and $M$ the 
canonical equilibrium magnetization. 
The equilibrium probability distributions 
found numerically are in fair agreement with these theoretical predictions.
However, around the critical energy,
relaxation to equilibrium depends in a very sensitive way on 
the initial conditions adopted.
When starting with ``water bag'' initial conditions, i.e.
 a flat probability distribution of finite width centered around zero 
for $g(p)$, and putting all particle positions $q_i$ at zero, we find 
quasi-stationary (long living) nonequilibrium states. 
These states have a lifetime which
increases with $N$, and are therefore stationary in the continuum
limit. 
We plot in fig.~1 the caloric curve
for these states in the case  N=20000. The points plotted are the result of an
integration of $0.5 \cdot 10^6$ time steps.
They are far from the equilibrium caloric curve 
around  $U_c$, showing a region of negative specific heat and a
continuation of the high temperature phase (linear $T$ vs. $U$ relation) into 
the low temperature one. 
It is very intriguing that this out-of-equilibrium
quasi-stationary states indicate a
caloric curve very similar to   that  one found
for first order phase transitions in Refs.~\cite{thir,compa,laba,gross,tor}.
 In that case,
however,  the corresponding states are stationary also at finite $N$. The 
coexistence of different states in the 
continuum limit near the critical region
is a purely microcanonical effect. It arises  after the inversion of the
$t \to \infty$ limit with the $N \to \infty$ one and could be
considered as the typical signature of critical slowing down.

We have studied how finite-size effects influence the behavior of the LLE.
In fig. 2(a) we plot $\lambda$ as a function of $U$ for various $N$
values. In the limit of very small and very large energies, the system is 
quasi-integrable, the Hamiltonian reducing to that of weakly coupled 
harmonic oscillators in the former case and to that of free rotators 
in the latter. 
In the region of weak chaos, for  $U < 0.25$, the curve has
a weak $N$-dependence. Then $\lambda$ changes abruptly and   
a region of strong chaos begins.
In Ref.~\cite{ruffo} it was observed that in between $U=0.2$
and $U=0.3$ 
a different dynamical regime sets in and particles start to evaporate 
from the main cluster. A similar regime was found in Ref.~\cite{cmd}. 
This behavior is  also similar to that found in Ref.~\cite{nayak} 
at the solid-liquid transition. 
In this region of strong chaoticity we observe a pronounced 
peak already for $N=100$~\cite{yama}, which persist and becomes broader
for $N=20000$. The location of the peak is just below the critical energy
at $U\sim 0.67$ and depends very weakly on N. 
At variance with what suggested in Ref.~\cite{cmd} 
the peak does not grow with $N$.

The standard deviation of the kinetic energy per particle 
$\sigma(K)/\sqrt{N}$ is plotted in fig. 2(b). In the low energy region 
this quantity is in agreement with the canonical  calculation 
$\sigma(K)/\sqrt{N}\propto U$. In correspondence to the Lyapunov peak, 
we observe also a sharp maximum of kinetic energy fluctuations, though
finite size effects are stronger for the LLE than for kinetic energy
fluctuations.  
Thus, probes of chaotic behavior (LLE) and thermodynamical quantities
(e.g. kinetic energy fluctuations) seem to be  strongly related.
We discuss here an intuitive interpretation of the relation
between LLE and kinetic fluctuations. Each of
the linearized equations (\ref{lin1}) contains diagonal (\ref{der1})
and off-diagonal (\ref{der2}) terms. Since the off-diagonal terms
result from a sum of incoherent terms, we can, in a first approximation,
neglect them. The diagonal term is of order $M^2$ 
(see Eqs. (\ref{der1}),(\ref{diag})) and averages to $<M^2> = T +1 -2U$. 
If this term would be constant in time, the LLE would be zero. In fact 
in this case one gets the equations for uncoupled harmonic oscillators.
However, there are fluctuations, which give a non-zero coupling,
whose standard deviation $\sigma (M^2)$ is related to the one of the
kinetic energy, $\sigma(M^2) = 2 \sigma(K)/{N}$, considering the 
relationship  $T=2 <K>/N$.
This indicates that the LLE is strictly related to 
kinetic energy fluctuations, but
this relation is not simple and quite difficult to extract analytically
(some indications in this sense were recently proposed also in 
Ref.~\cite{aldoprep}).
In the low energy phase ($U<0.25$) it is possible to work out a more
stringent relation. In this case, the components of the tangent vector 
sum up incoherently to give for the average growth a term of the size 
$\sqrt{N} M^2$. It is then quite natural to associate the Lyapunov exponent 
to the inverse time-scale given by the fluctuations of the average growth
\begin{equation}
\lambda^2 \sim \sigma (\sqrt{N} M^2) \sim 2 \frac{\sigma(K)}{\sqrt{N}}~,
\label{stima}
\end{equation}
Then, substituting the canonical estimation for kinetic energy fluctuations in 
Eq.~(\ref{stima}), we get $\lambda \propto \sqrt{U}$. We have tested 
numerically this prediction (see fig.3(a)). The $1/2$ power law at small 
$U$-values is fully confirmed.
We have checked numerically that off-diagonal
terms (\ref{diag}) cannot be completely neglected 
- expecially in the strong chaotic region. 
This latter important remark is also relevant for the application of a 
recently derived formula for the LLE~\cite{lapo1} 
(see also~\cite{latora,firpo}).
A similar power law behavior was also found for other systems~\cite{nota1}.

At variance with the N-independent behavior observed at small energy $U$, 
strong finite size effects are present above $U_c$.
In fig.3(b) we show, for $U > U_c$, how the LLE goes to zero as a 
function of $N$. We also plot in the same figure a calculation of the LLE 
using a random distribution of particle positions $q_i$ on the circle
in the equations for the tangent vector (\ref{lin1}).
The agreement between the deterministic estimate and this random matrix
calculation is very good.
We find also that $\lambda$ scales as $N^{- {1\over3}}$, as indicated  
by the fit in fig.~3(b).
This can be explained by means of an analytical result obtained for the 
LLE of product of random matrices~\cite{pari}.
If the elements of the symplectic random matrix have zero mean, the LLE 
scales with the power $2/3$ of the perturbation. In our case, the latter 
condition is satisfied and the perturbation is the magnetization $M$. 
Since $M$ scales as $N^{-{1\over 2}}$, we get the right scaling of 
$\lambda$ with $N$. This proves that the system is integrable
for $U \ge U_c $ as $N \to \infty$. 
This result is also confirmed by a recent more sophisticated theoretical 
calculation~\cite{firpo}.

%%%%%%%%%%%%%%%%%%%%%%%%%%%%%%%%%%%%%%%%%%%%%%%
%%%  Conclusioni

In conclusion, we have investigated the Lyapunov instability for a system with 
long-range forces showing a second order phase transition.  
We found strong finite size effects in the LLE.
The LLE is peaked just below the critical energy, where kinetic fluctuations 
are maximal. Away from the transition region, the LLE goes to zero with 
universal scaling laws which can be explained by simple theoretical arguments.
We think that this toy model contains all the main ingredients to understand
the general  behavior of the LLE in more realistic situations.   
%%%%%%%%%%%%%%%%%%%%%%%%%%%%%%%%%%%%%%%%%%%%%%%%%%
   
We thank M. Baldo, A. Bonasera, F. Leyvraz, and M. Ploszajczac 
for useful discussions. A special thank goes to M. Antoni for having 
insisted to use equilibrated Gaussian initial states and to A. Torcini
for several crucial suggestions. V.L. and S.R. thank INFN for financial support.
S.R. thanks CIC, Cuernavaca, Mexico for financial 
support. This work is also part of the European contract No. ERBCHRXCT940460 
on "Stability and universality in classical mechanics".

$(a)$ E-mail: latora@ctp.mit.edu

$(b)$ E-mail: andrea.rapisarda@ct.infn.it 

$(c)$ Permanent address, Dipartimento di Energetica, Universit\'a di
Firenze, Via S. Marta, 3 I50139, Firenze, Italy, INFN,
Firenze, E-mail: ruffo@ing.unifi.it

{\bf FIGURE CAPTIONS}
%%%%%%%%%%%%%%%%%%%%%%%%

%\bigskip
Fig.1~~~     Theoretical predictions 
             in the canonical ensemble  
             (full curve) for T vs. U in comparison 
             with numerical simulations 
             (microcanonical ensemble) for 
             N=100,1000, 5000, 20000. The 
             vertical line indicates the canonical 
             critical energy $U_c=0.75$. 
             We plot also the microcanonical results for 
              the Quasi-Stationary States (QSS) in the case
             N=20000 (full circles).
             In the 
             inset we show the magnetization
             vs. $U$, again the full line is 
             the canonical theoretical prediction. 
             
%%%%%%%%%%%%%%%%%%%%%%%%
%
%\bigskip

Fig.2~~~     (a) Numerical calculation of the 
             largest Lyapunov exponent as 
             a function of $U$ for various system 
             sizes: N=100,1000, 5000 
             and 20000. (b) Kinetic energy 
             fluctuations vs. $U$.
             The vertical line indicates the 
             canonical critical energy $U_c=0.75$ 

%\bigskip

Fig.3~~~     Behavior of  the largest Lyapunov exponent 
             (LLE) for $U$ much 
             smaller (a) and much greater (b) than $U_c=0.75$. 
             In panel (a) LLE shows a 
             universal law which can be fitted by a 1/2 power law 
             (full line). No N-dependence is found. 
             In (b)  the LLE - for different N and 
             energies - is compared with a calculation
             done  with a random choice 
             of particle positions (diamonds). 
             The latter follow a  power law 
             with an exponent -1/3 (dashed line)
             (see text).
%
%%%%%%%%%%%%%%%%%%%%%%%%%%%%%%%%%%%%%%%%%%%%%%%


\begin{thebibliography}{99}
\bibitem{nota1}
Actually, in refs. \cite{cmd,aldoprep} this behaviour was found as a 
function of temperature. Anyway, in our case $T \sim U$ at
small energies (a similar relation exists at high energy
$T=2U+1$).
\end{thebibliography}
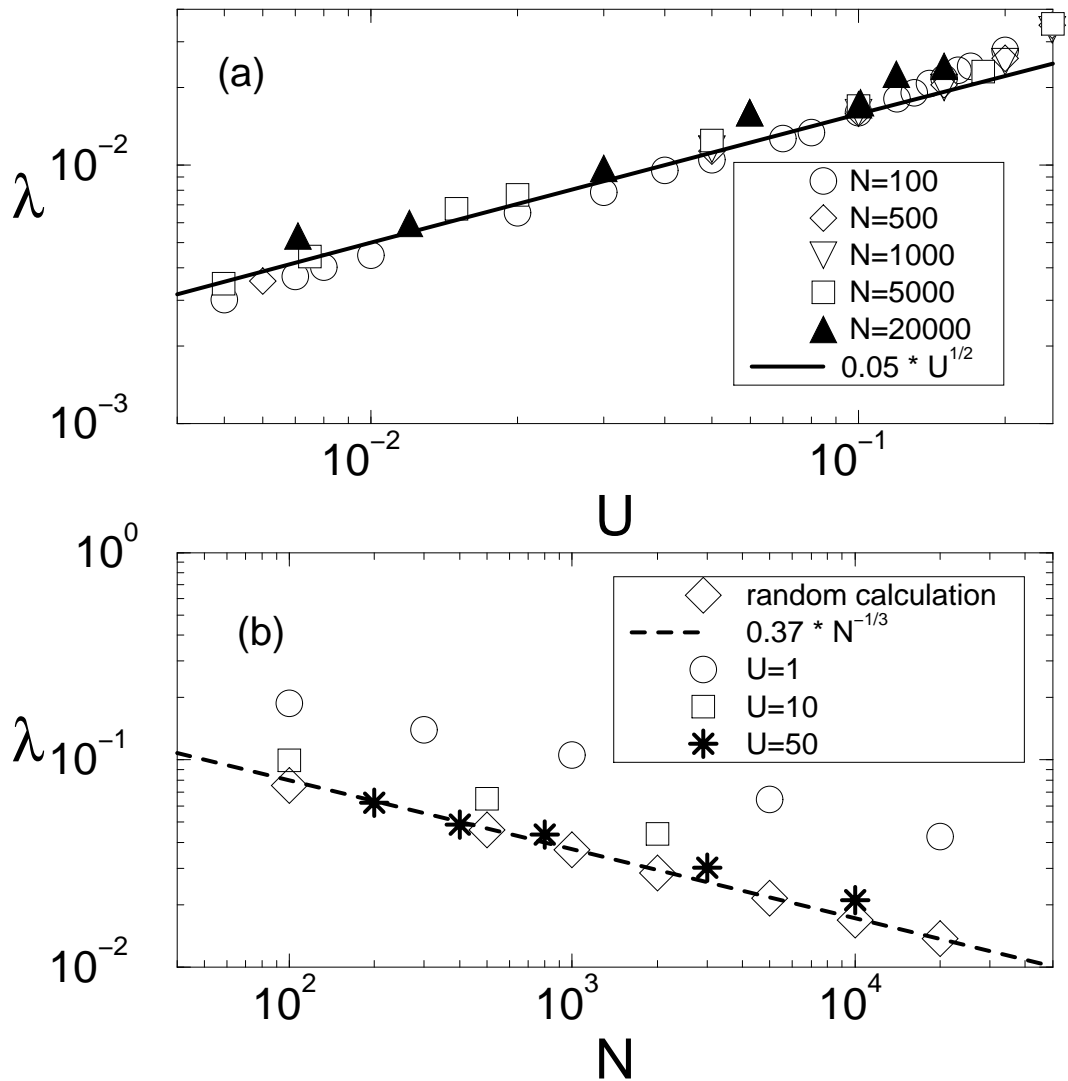
\end{document}